\documentclass[%
 reprint,
superscriptaddress,
amsmath,amssymb,
aps,prb
]{revtex4-2}
\usepackage{graphicx}
\usepackage{dcolumn}
\usepackage{bm}%
\usepackage{graphicx,graphics,color,epsfig}
\usepackage{amsmath}
\usepackage{mathrsfs}
\usepackage{amssymb}
\usepackage{subfigure}
\usepackage{amsfonts}
\usepackage{graphicx}
\usepackage{multirow,booktabs}
\usepackage{float}
\usepackage[normalem]{ulem}  
\usepackage{xcolor}
\begin{document}

\title{Loschmidt echo zeros in finite-size quantum systems with linear quench}
\author{Zhen-Yu Zheng}
\affiliation{Beijing National Laboratory for Condensed Matter Physics, Institute of Physics, Chinese Academy of Sciences, Beijing 100190, China}

\author{Xudong Liu}
\affiliation{Beijing National Laboratory for Condensed Matter Physics, Institute of Physics, Chinese Academy of Sciences, Beijing 100190, China}
\affiliation{School of Physical Sciences, University of Chinese Academy of Sciences, Beijing 100049, China}

\author{Siyan Lin}
\affiliation{Beijing National Laboratory for Condensed Matter Physics, Institute of Physics, Chinese Academy of Sciences, Beijing 100190, China}
\affiliation{School of Physical Sciences, University of Chinese Academy of Sciences, Beijing 100049, China}

\author{Yu Zhang}
\affiliation{Beijing National Laboratory for Condensed Matter Physics, Institute of Physics, Chinese Academy of Sciences, Beijing 100190, China}
\affiliation{School of Physical Sciences, University of Chinese Academy of Sciences, Beijing 100049, China}

\author{Shu Chen}
\email{schen@iphy.ac.cn }
\affiliation{Beijing National Laboratory for Condensed Matter Physics, Institute of Physics, Chinese Academy of Sciences, Beijing 100190, China}
\affiliation{School of Physical Sciences, University of Chinese Academy of Sciences, Beijing 100049, China}

\date{\today}

\begin{abstract}
Dynamical quantum phase transitions reveal singularities in quench dynamics, characterized by the emergence of Loschmidt echo zeros at critical times, which usually exist only in the thermodynamic limit but are absent in finite-size quantum systems. In this work, we propose a theoretical scheme to probe Loschmidt echo zeros in finite-size systems by applying a two-step quenching protocol, which offers an experimentally feasible approach to study Loschmidt echo zeros.
Using the transverse Ising model as a testbed, we identify that the exact Loschmidt echo zeros can be always accessed by tuning the quench rate, when the quench is across the phase transition point. The associated rate function displays divergence at critical times, accompanying with the change of the dynamical topological order parameter. The critical times are influenced by the quench rate, system size, and momentum modes, embodying the interplay between finite-size effects and critical dynamics. Moreover, the generality of these observations is further confirmed in the XY and Haldane models.
\end{abstract}

\maketitle

{\it Introduction.-}
Dynamical quantum phase transition (DQPT) \cite{heyl2013prl,heyl2018rpp,heyl2019el} has emerged as an important concept for understanding singular dynamical behaviors in quantum systems out of equilibrium.
DQPTs can be identified through the emergence of Loschmidt echo zeros (LEZs), which play a role analogous to the Lee-Yang or Fisher zeros of the partition function in statistical mechanics \cite{heyl2013prl,Lee-Yang1,Lee-Yang2,Fisher}. The corresponding rate function \cite{heyl2013prl,heyl2015prl} and dynamical topological order parameters (DTOPs) \cite{heyl2016prb} exhibit non-analyticities at critical times, marking the occurrence of a dynamical phase transition.  With the rapid development of quantum simulation technologies, the study of non-equilibrium dynamics in isolated quantum systems has garnered significant attention \cite{Bloch2008rmp,Nori2014rmp}. Signature of DQPTs has been experimentally observed in platforms ranging from cold atomic systems, trapped ions and  quantum simulator platforms \cite{Jurcevic2017prl,Weitenberg2018NP,Monroe2017N,Fan2019PRAp,Thywissen2019SA,xue2019prl,Li2020prl,Duan2020prl}.
While most studies of DQPTs focused on sudden quenches \cite{heyl2013prl,Heyl2014,Karrasch2013PRB,Kriel2014PRB,Sirker2014prb,Garrahan2014prb,Dora2014prb,Kehrein2015prb,Vajna2015prb,Dutta2016pre,YangC,Halimeh,Sharma,Dutta2017prb,Sacha2018pra,
Fan2018prb,Bera2019prb,ZhouBZ2019,Khatun2019prl,Puebla2020prbr,SunG,Yang2023prb,Rubtsov2023prb,Yu2024prb,Zhou2021prb,Zeng2023prb}, slow quenches have been studied mainly in the context of Kibble-Zurek (KZ) scaling of the defect density and the residual energy
\cite{Damski2005prl,Zurek2005prl,Dziarmaga2005prl,Polkovnikov2005prb,Dziarmaga2010ap} and also in the situation of DQPTs \cite{Dutta2016prb,Langari2024jpcm,Schuricht2016scipost,Tong2024pra}.

Generally, exact LEZs are accessible only  in the thermodynamical limit. As actual experimental systems are always limited in size, it is important to understand why and how non-analyticities can emerge in finite-size systems. In order to access LEZs, usually one needs to make analytic continuation of the time or phase driving parameters to the complex plane. These Fisher or Lee-Yang zeros localize on the real time axis only when the lattice size tends to infinite \cite{heyl2013prl,Dutta2016prb}. Alternative schemes to access LEZs of finite-size systems are proposed in Refs.\cite{Zhou2021prb,Zeng2023prb} in the framework of sudden quench. These schemes either rely on fine tuning of quench parameters or need to introduce an additional twist boundary parameter, making it difficult for the experimental detection of LEZs.

In this work, we propose a new theoretical scheme to detect LEZs in finite-size systems by applying a two-step quenching protocol, which offers an experimentally feasible approach to detect LEZs. The protocol consists of a linear quench to the final system parameters, followed by evolution under the final Hamiltonian. We take the quantum Ising model as a concrete example to illustrate our idea and results. By analyzing  the condition for the occurrence of exact LEZs, we identify that LEZs can be always accessed by tuning the quench rate to a series of discrete values when the initial and final transverse fields belong to distinct quantum phases. Whenever the LEZs occur at critical times, the associated rate function is divergent. We also show that the DTOP changes its integer value at these critical times when the system size is large.
Our results demonstrate that the critical times are influenced by the quench rate and system size, highlighting the interplay between finite-size effects and critical dynamics. Our work clarifies the mechanism for the occurrence of LEZs in realistic finite-size systems and establishes a practical framework for probing LEZs and the divergence of rate function in experimental settings with no need to introduce any non-physical complex parameter.



{\it Model and scheme.-}
We consider the quantum Ising model with a transverse field  defined as
\begin{equation}\label{model}
H=\sum_{j=1}^{N}-(J \sigma_{j}^{x}\sigma_{j+1}^{x}+h \sigma_{j}^{z}),
\end{equation}
where $\sigma_{j}^{\alpha}$ are Pauli matrices and the periodic boundary condition is applied. For convenience, we assume the lattice number $N$ is even and set $J=1$ as the unit of energy. By using the Jordan-Wigner transformation and the Fourier transform, we get the Hamiltonian in the momentum space given by \cite{Dziarmaga2005prl,SM}:
\begin{equation}
H =2 \sum_{k>0} \left(\begin{matrix} c_{k}^{\dagger},  c_{-k} \end{matrix}\right)
\left(\begin{matrix} h-\cos{k} & \sin{k} \\
\sin{k} & \cos{k}-h \end{matrix}\right)
\left(\begin{matrix} c_{k} \\ c_{-k}^{\dagger} \end{matrix}\right).
\end{equation}
Since the initial state is taken as the ground state of the pre-quench system, we only need consider the subspace of even parity with the corresponding pseudo-momenta $k \in [\pm\frac{\pi}{N}, \pm\frac{3\pi}{N}, \ldots, \pm \frac{(N-1) \pi}{N}]$.
By applying the Bogoliubov transformation
$
c_{k}=u_{k}^* \gamma_{k}+v_{k} \gamma_{-k}^{\dagger},
$
the Hamiltonian is diagonalized as
\begin{equation}
H=\sum_{k} \epsilon_{k}\left(\gamma_k^{\dagger} \gamma_k-\frac{1}{2}\right),
\end{equation}
where
$
\epsilon_{k}  = 2 \sqrt{(h-\cos{k})^2+(\sin{k})^2}
$,
$u_k  = \frac{\epsilon_k/2+h-\cos{k}}{\sqrt{\epsilon_k(\epsilon_k/2+h-\cos{k})}}$, and
$ v_k  = -\frac{\sin{k}}{\sqrt{\epsilon_k(\epsilon_k/2+h-\cos{k})}}$.

\begin{figure}[!h]
  \centering
  \includegraphics[width=8.25cm]{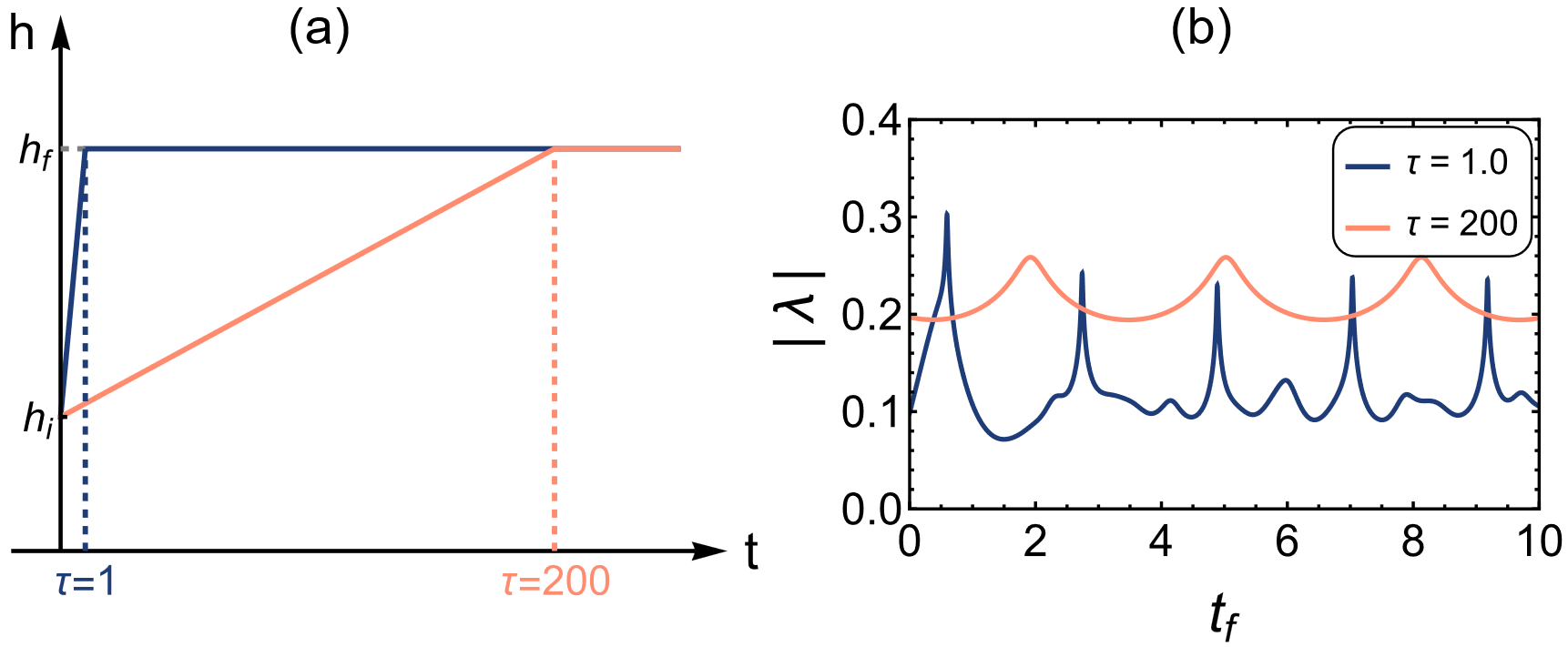}
  \caption{(a) Illustration of the time-dependent transverse field $h_{t}$ during and after the linear quench. The field linearly varies from $h_{i}$ to $h_{f}$ over the quench duration $\tau$, and remains constant at $h_{f}$ afterward.(b) Rate function $|\lambda|$ versus $t_f$ for system with $N=50$ and the quench time $\tau=1$ and $200$, respectively.}
  \label{figsyt}
\end{figure}

We consider a two-step time-evolution process as illustrated in Fig.\ref{figsyt} (a). In the first step, we control the strength of the transverse field $h$ to vary linearly. At $t=0$, the strength is assumed to be $h_{i}$ and after a time $\tau$, it linearly changes to $h_{f}$. In the second step, after the linear ramp quench, the transverse field strength is fixed at $h_{f}$ and the state evolves under the final Hamiltonian $h_{f}$. For convenience, we assume the strength of the transverse field  $h_{i}>0$, $h_{f}>0$ and the time dependence of the transverse field strength is given by:
\begin{equation}
h(t) = \begin{cases}
            h_{i}+ (h_{f}-h_{i})t/\tau  & t \leq \tau, \\[8pt]
            h_{f} & t > \tau .
          \end{cases}
\end{equation}
The quench rate of the linear ramp is $v=(h_{f}-h_{i})/{\tau}$. When $\tau\rightarrow0$, the quench process reduces to a sudden quench.


We firstly focus on the step involving the linear ramp quench ($t\leq \tau$). The initial state is chosen to be the ground state of the initial Hamiltonian with the transverse field strength $h_{i}$. 
Explicitly, the initial state is given by
$
|\varphi_{i}\rangle=\prod_{k>0}(u_{k,i}+v_{k,i}c_{k}^{\dag} c_{-k}^{\dag})|0\rangle\propto \prod_{k>0}\gamma_{k}\gamma_{-k}|0\rangle
$,
where $\epsilon_{k,i}=\epsilon_{k}|_{h=h_i}, u_{k,i}=u_k|_{h=h_i}, v_{k,i}=v_{k}|_{h=h_i}$.
Then, we work in the Heisenberg picture, defining the annihilation operator $c_{k,t}$ by a time-dependent Bogoliubov transformation as
$
c_{k,t}=u_{k,t}^* \gamma_k+v_{k,t}\gamma_{-k}^{\dagger}
$.
The Heisenberg equation is equivalent to the dynamical version of the Bogoliubov-de Gennes equation:
\begin{equation}
\begin{aligned}
& i \frac{d}{d t} u_{k,t}= 2 u_{k,t} [\cos{k}-h_{t}] +2 v_{k,t} \sin{k} , \\
& i \frac{d}{d t} v_{k,t}= 2 v_{k,t} [h_{t}-\cos{k}] +2 u_{k,t} \sin{k}.
\end{aligned}\label{tbdg}
\end{equation}
For convenience, we assume $\hbar=1$. This problem can be mapped to a two level Landau-Zener problem. The instantaneous state after the linear ramp quench step is
\begin{equation}
|\varphi_{\tau}\rangle=
\prod_{k>0}(u_{k,\tau}+v_{k,\tau}c_k^\dag c_{-k}^\dag)|0\rangle,
\end{equation}
where $u_{k,\tau}$ and $v_{k,\tau}$ are the solutions to the partial differential equations Eq.(\ref{tbdg}) with the initial condition $u_{k,0}= u_{k,i}, v_{k,0}=v_{k,i}$.

We further proceed to the step with the state evolving under the final Hamiltonian ($t\geq \tau$). The final state can be expressed as
\begin{equation}
|\varphi_{f}\rangle = e^{-i H_{f} t_{f}} |\varphi_{\tau}\rangle,
\end{equation}
where the time $t_{f}$  is defined as the difference between the total time and the duration of the linear quench, i.e., $t_{f}=t-\tau$.
It follows that the Loschmidt amplitude is given by
\begin{equation}
\mathcal{G}(t_{f})=\langle \varphi_{i} |\varphi_{f}\rangle=\langle \varphi_{i} | e^{-i H_{f} t_{f}} |\varphi_{\tau}\rangle,
\end{equation}
and the Loschmidt rate function is
\begin{equation}
\lambda= -\frac{1}{N} \ln\left[|\mathcal{G}(t_{f})|^{2}\right].
\end{equation}
Before carrying out further analysis, we would like to indicate the difference of our motivation from that in some previous references \cite{Dutta2016prb,Schuricht2016scipost,Tong2024pra}. While Ref. \cite{Tong2024pra} focused on DQPTs occurring during the ramp process, authors of Ref. \cite{Dutta2016prb} studied the evolution after the quench but considered the finite-time ramp quench as a way to prepare the ``initial state''. Ref. \cite{Schuricht2016scipost} considers the full-time evolution but restricts its analysis to the thermodynamic limit. In contrast, our work focuses on the explicit identification of exact LEZs in finite-size systems. By tuning the quench rate, we gain direct access to the strict LEZs.

In terms of the spectrum decomposition, the final Hamiltonian $H_{f}$ can be expressed as:
$
H_{f}=\sum_{k}\epsilon_{k,f} (|\varphi_{k,f}^{+}\rangle\langle\varphi_{k,f}^{+}|-|\varphi_{k,f}^{-}\rangle\langle\varphi_{k,f}^{-}|)
$,
where the energy is $\epsilon_{k,f} = \epsilon_{k}|_{h=h_f}$, the positive energy state is $|\varphi_{k,f}^{+}\rangle = (u_{k,f}, - v_{k,f})^{\mathrm{T}}$ and the negative energy state is $|\varphi_{k,f}^{-}\rangle = (v_{k,f}, u_{k,f})^{\mathrm{T}}$. Then, the Loschmidt amplitude can be simplified as:
\begin{equation}
\mathcal{G}(t_{f}) = \prod_{k>0}(A_{k,\tau,i,f}e^{i \varepsilon_{k,f} t_{f}}+ B_{k,\tau,i,f} e^{-i \varepsilon_{k,f} t_{f}}),
\label{GN1}
\end{equation}
where the coefficients $A_{k,\tau,i,f}$ and $B_{k,\tau,i,f}$ are
\begin{equation*}
\begin{aligned}
A_{k,\tau,i,f}&=(u_{k,i}u_{k,f}+v_{k,i}v_{k,f})(u_{k,\tau}u_{k,f}+v_{k,\tau}v_{k,f}),\\
B_{k,\tau,i,f}&=(u_{k,i}v_{k,f}-v_{k,i}u_{k,f})(u_{k,\tau}v_{k,f}-v_{k,\tau}u_{k,f}),
\end{aligned}
\end{equation*}
and $u_{k,i}, v_{k,i}, u_{k,f}, v_{k,f}$ can be chosen to be real. By using Eq.(\ref{GN1}), we can calculate the Loschmidt amplitude and the Loschmidt rate function $\lambda$. An illustration of $|\lambda|$ versus $t_f$ is presented in Fig.\ref{figsyt} (b). While the rate function displays a series of peaks for $\tau=1$, no obvious peaks can be observed for a slow ramp with $\tau=200$.

{\it Exact Loschmidt echo zeros.-}
Now we explore the condition for the occurrence of LEZs, i.e., $\mathcal{G}(t)=0$, which gives the following constraint condition:
\begin{equation}\label{cba}
\left|A_{k,\tau,i,f}\right| = \left|B_{k,\tau,i,f}\right|.
\end{equation}
When the above relation is fulfilled, a series of critical times for the occurrence of divergence of rate function are given by:
\begin{equation}\label{ct}
t_{c}=\frac{\pi}{\varepsilon_{k,f}}\left(n+\frac{1}{2}\right)-\frac{\phi_{k,\tau,i,f}}{2\varepsilon_{k,f}},
\end{equation}
where $\phi_{k,\tau,i,f}=\mathrm{Arg}\left[\frac{A_{k,\tau,i,f}}{B_{k,\tau,i,f}}\right]$.

Consider the limit case with $\tau \rightarrow 0$, which reduces back to the sudden quench with the quench rate $v\rightarrow\infty$. In this case, we have
$u_{k,\tau}\rightarrow u_{k,i}$ and $v_{k,\tau}\rightarrow v_{k,i}$. Substituting these into Eq.~(\ref{cba}), we see that only when the $k$-mode satisfies the condition
\begin{equation}
k_{s} = 2 \text{arctan}\left[\frac{\sqrt{-(h_{i}-1)(h_{f}-1)}}{\sqrt{(1+h_{i})(1+h_{f})}}\right],
\label{ks}
\end{equation}
exact LEZs can occur.
The real-valued momentum $k_{s}$ imposes constraints on $h_{i}$ and $h_{f}$: $0 < h_{i} < 1$, $h_{f} > 1$, or $0 < h_{f} < 1$, $h_{i} > 1$. For the transverse field Ising model, $h/J=1$ marks the phase transition point. Therefore, only when the initial and final field strengths lie in distinct phase region, $|A_{k,\tau,i,f}|$ and $|B_{k,\tau,i,f}|$ can be equal, and divergence of rate function can occur. Nevertheless, for given $h_i$ and $h_f$, $k_{s}$ determined by Eq.(\ref{ks}) does not usually happen to taking the discrete values $\pm\frac{\pi}{N}, \pm\frac{3\pi}{N}, \ldots, \pm \frac{(N-1) \pi}{N}$, but can be approached in the thermodynamic limit $N \rightarrow \infty$. This explains why exact LEZs usually only occur in the thermodynamic limit.

\begin{figure}[t]
\centering
\includegraphics[width=8cm]{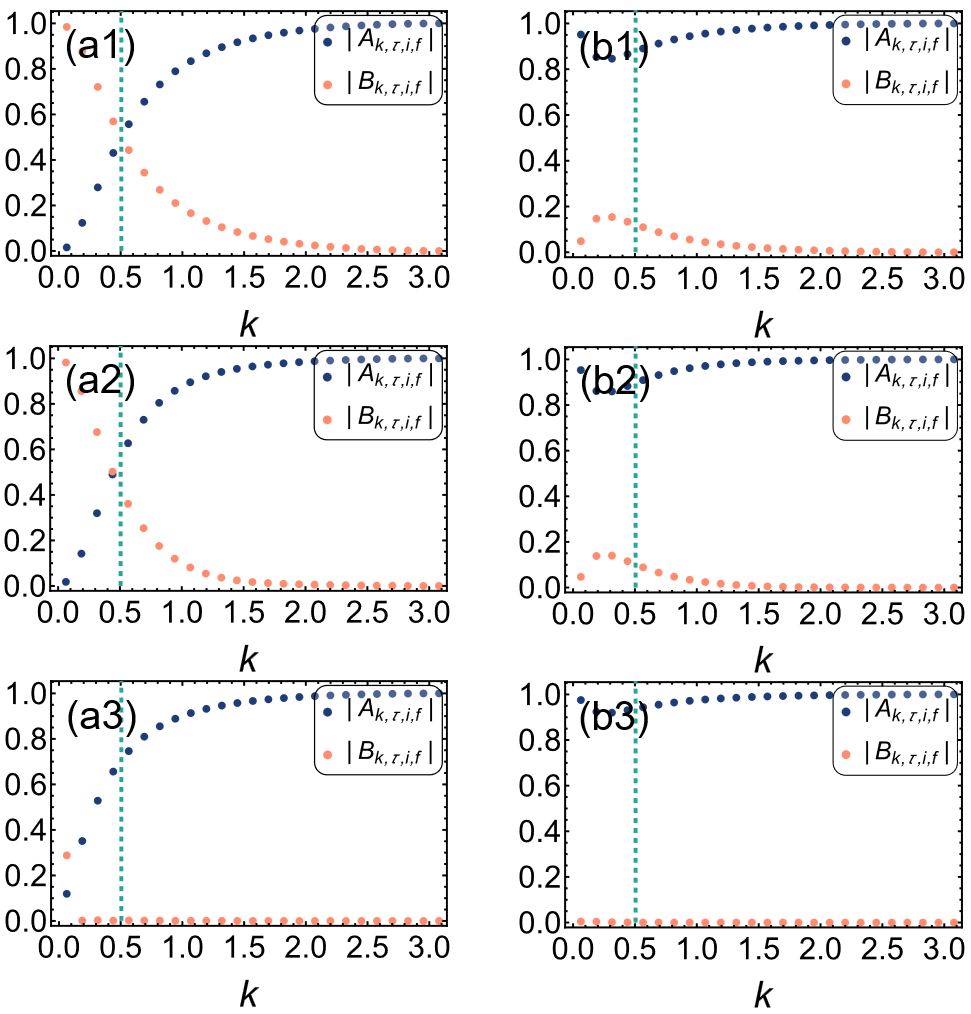}
\caption{Panels (a1)-(a3) and (b1)-(b3) show the behavior of the complex coefficients $|A_{k,\tau,i,f}|$ (blue dots) and $|B_{k,\tau,i,f}|$ (orange dots) under varying quench times with $N=50$. The quench time is set to $\tau=0.01$ in (a1), (b1), $\tau=1$ in (a2), (b2), and $\tau=100$ in (a3), (b3). The initial transverse field strength is $h_{i}=0.5$, while the final transverse field strength is $h_{f}=1.5$ for (a1)-(a3) and $h_{f}=0.9$ for (b1)-(b3). The vertical green dashed line indicates the critical $k_{s}$.}
\label{figAB}
\end{figure}

Now we consider the general case with finite quench rate $v$. In this case, the quench rate can be used as a tuning parameter, so that Eq.~(\ref{cba}) can be fulfilled even for discrete $k$ modes by tuning $v$. 
To get a straightforward understanding, we numerically calculate $|A_{k,\tau,i,f}|$ and $|B_{k,\tau,i,f}|$ with discrete values of $k$ for various $\tau$, $h_i$ and $h_f$. In Fig.\ref{figAB}, we analyze the coefficients $|A_{k,\tau,i,f}|$ and $|B_{k,\tau,i,f}|$ under varying quench times. In the sudden-quench limit, the largest $k$-mode $k_s$ is realized.  As shown in Fig.\ref{figAB}(a1)-(a3), for a quench across different phases, the $k$-mode satisfying $|A_{k,\tau,i,f}|=|B_{k,\tau,i,f}|$ shifts toward smaller momenta as $\tau$ increases ($k<k_s$) .  Upon increasing $\tau$, a maximal value $\tau_{\max}$ is reached, set by the smallest momentum mode. Beyond this point no $k$-mode fulfills the condition, and the dynamics becomes adiabatic, thereby precluding LEZs. In contrast, for a quench within the same phase, no momentum mode satisfies $|A_{k,\tau,i,f}|=|B_{k,\tau,i,f}|$ for any $\tau$, as illustrated in Fig.\ref{figAB}(b1)-(b3). Hence, intra-phase quench does not induce divergence of rate function in the quantum Ising model.


\begin{figure}[t]
\centering
\includegraphics[width=8.25cm]{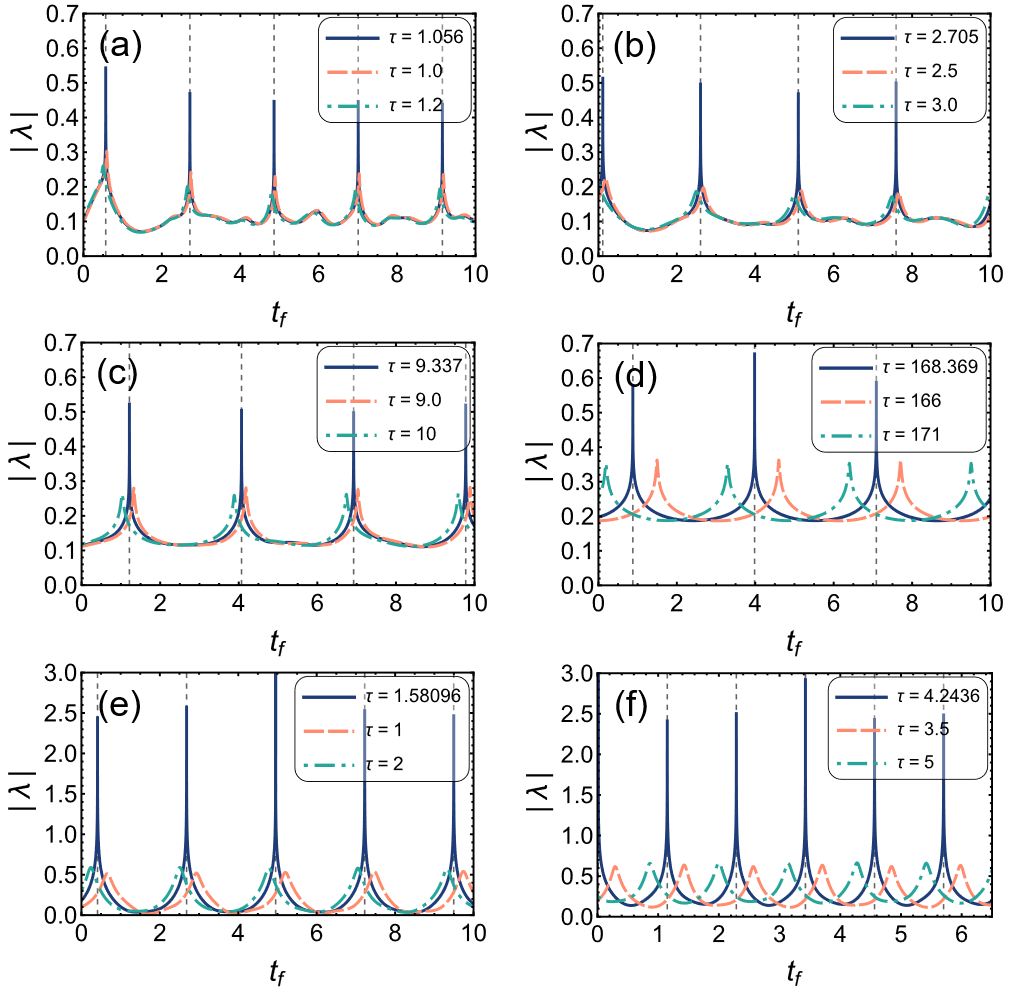}
\caption{Time evolution of the rate function $\lambda$ for the quantum Ising model for different values of $k$ , highlighting the non-analytic line (gray dashed line) predicted by Eq.(\ref{ct}). Panels (a)-(d) correspond to $k = \frac{7\pi}{50}, \frac{5\pi}{50}, \frac{3\pi}{50} ,$ and $\frac{\pi}{50} ,$ with $N=50$. Panels (e)-(f) correspond to $k=\frac{\pi}{8}$ with $N=8$.}
\label{figisingrf}
\end{figure}


For $h_{i}=0.5$ and $h_{f}=1.5$, we have $k_{s}=0.50536$ according to Eq.(\ref{ks}). Consider a finite-size system with $N=50$. The allowed momenta, which fulfill $k<k_s$, are $k\in \{\frac{7\pi}{50}, \frac{5\pi}{50}, \frac{3\pi}{50}, \frac{\pi}{50}\}$. Substituting these momenta into Eq.(\ref{cba}), we can find the corresponding values of $\tau$ given by $\tau_c=1.056$, $2.705$, $9.337$ and $168.369$, respectively. The value of $\tau_c$ increases with the decrease of $k$ and reaches the maximum at $k=\frac{\pi}{N}$.  As shown in Fig.\ref{figisingrf}(a)-(d), when $\tau$ takes these critical values, exact LEZs occur periodically at  a series of critical times,  witnessed by the divergence of the Loschmidt rate function at the predicted critical times from Eq.(\ref{ct}). When $\tau$ deviates from the critical value $\tau_c$, no divergence can be observed, but a series of obvious peaks are still observed when the deviation of $\tau_c$ is small. If the deviation is large, for example, $\tau=200$ as displayed in Fig.\ref{figsyt}(b), no obvious peaks can be observed.  We also note that the period for the occurrence of divergence of rate function increases with decreasing quench rate, as suggested by Eq.(\ref{ct}). Slower quenching (larger $\tau_c$) reduces the value of $k$, lowering $\varepsilon_{k,f}$ and extending the period. Moreover, this phenomenon persists even in a small system of size $N=8$, as show in Fig.\ref{figisingrf}(e) with $h_{i}=0.5$ and $h_{f}=1.5$ and (f) with $h_{i}=0.25$ and $h_{f}=2.25$. For further illustration, we provide in the Supplemental Material a detailed analysis of the DTOP properties of this model, which demonstrates a direct correspondence with the observed LEZs \cite{SM}.

\begin{figure}[!h]
  \centering
  \includegraphics[width=8.25cm]{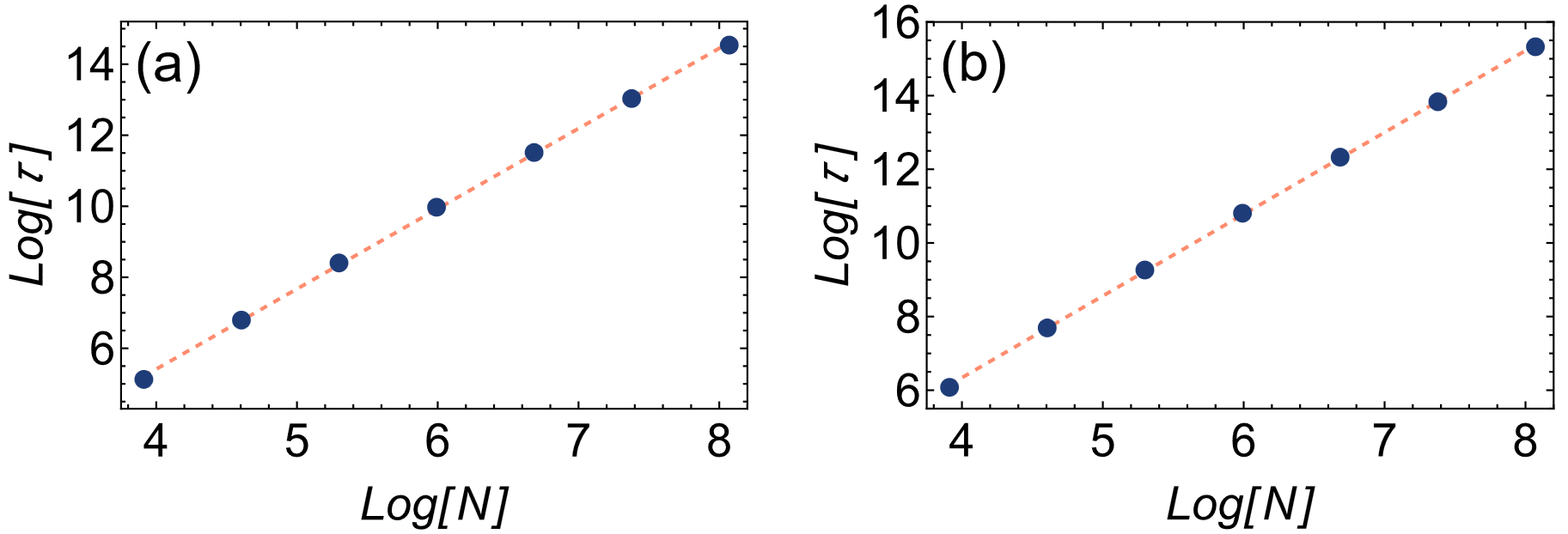}
  \caption{Finite size scaling of the quench time $\tau$ and system size $N$ with (a) $h_{i}=0.5, h_{f}=1.5$ and (b) $h_{i}=0.25, h_{f}=2.25$. The blue points are the data by solving Eq.(\ref{cba}) and the orange
  dashed lines are the fit data.}
  \label{figisingsc}
\end{figure}

Based on the above analysis, we can determine the slowest critical rate (or equivalently, the maximum critical quench time $\tau$) at which divergence of rate function occurs. This rate is governed by the smallest allowed momentum mode, corresponding to $k=\frac{\pi}{N}$. Therefore, the maximum quench time $\tau$ is directly related to the lattice size $N$. In Fig.\ref{figisingsc}, we illustrate the finite-size scaling of the maximum quench time, which is well fitted by $\tau= 0.02689 N^{2.2579}$ for Fig.\ref{figisingsc}(a) and $\tau= 0.07878 N^{2.2203}$ for Fig. \ref{figisingsc}(b). When $\tau$ becomes sufficiently large, surpassing the characteristic time scale set by the lowest $k$-mode that governs the occurrence of LEZs, the system undergoes an adiabatic evolution. Consequently, no divergence of rate function can be observed in this regime. We also investigate the scaling behaviors observed in the XY and Haldane models \cite{SM}. However, whether a universal mechanism governs this scaling behavior remains an open question.

{\it LEZs in another models.-}
To demonstrate the broad applicability of our theoretical framework, we extend the analysis to other typical quantum systems exhibiting DQPTs. As a representative example, the Hamiltonian of the XY model is given by
\begin{equation}
H=-\sum_{j=1}^N \left[\frac{1+\kappa}{2}\,\sigma_j^x\sigma_{j+1}^x + \frac{1-\kappa}{2}\,\sigma_j^y\sigma_{j+1}^y + h\,\sigma_j^z \right].
\end{equation}
The control parameters governing the phase transition are the anisotropy parameter $\kappa$ and the transverse field $h$.

\begin{figure}[t]
\centering
\includegraphics[width=8.25cm]{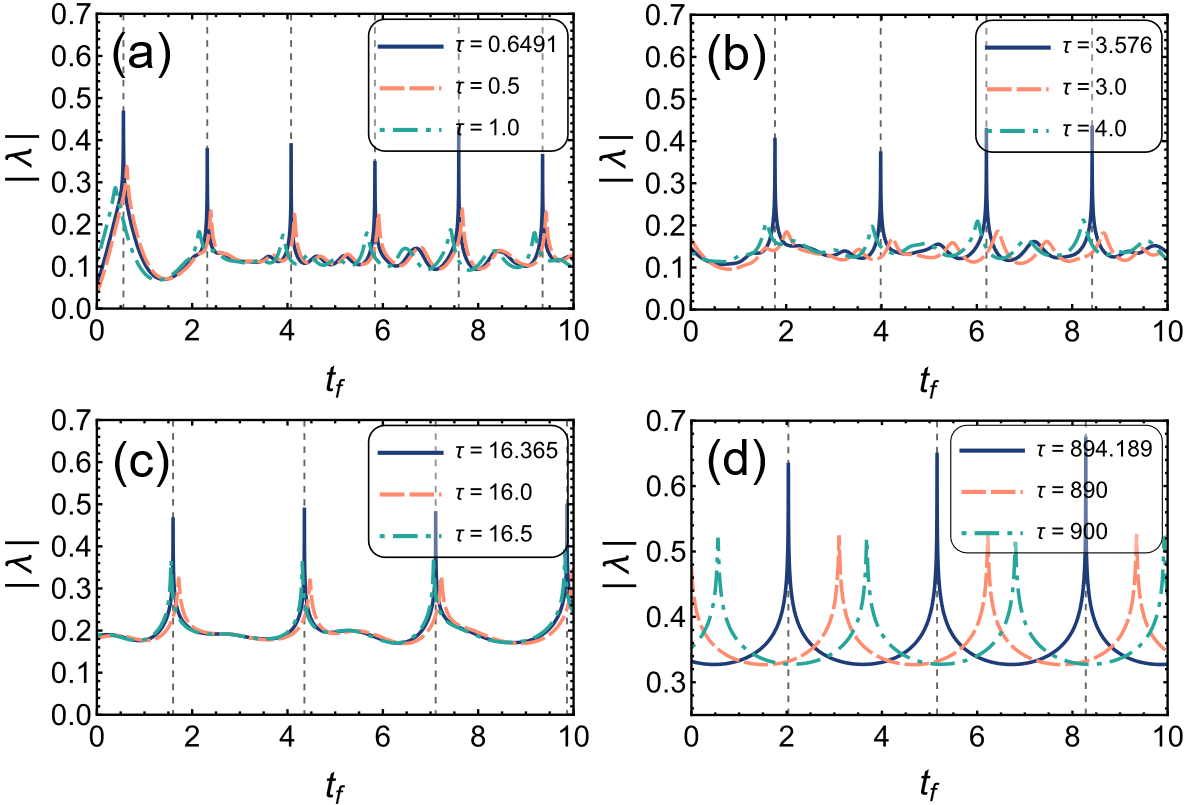}
\caption{Time evolution of the rate function $ \lambda $ for the XY model for different values of $ k $, with the non-analytic line (gray dashed) predicted by Eq.(\ref{ct}). Panels (a)-(d) show results for $ k = \frac{13\pi}{50}, \frac{9\pi}{50}, \frac{5\pi}{50}, $ and $ \frac{\pi}{50} $, respectively.}
\label{figxy}
\end{figure}

To observe LEZs, we analyze the non-analytic behavior of the Loschmidt rate function in a finite-size XY chain, as shown in Fig.~\ref{figxy}. For fixed $\kappa_i=\kappa_f=0.5$, we consider a quench from $h_i=0.5$ to $h_f=1.5$, corresponding to different equilibrium phases. The LEZ condition is satisfied for momenta
$k \in \{\tfrac{13\pi}{50}, \tfrac{11\pi}{50}, \tfrac{9\pi}{50}, \tfrac{7\pi}{50}, \tfrac{5\pi}{50}, \tfrac{3\pi}{50}, \tfrac{\pi}{50}\}$,
and representative cases are shown in Fig.\ref{figxy}(a)-(d), where the Loschmidt rate function exhibits strict non-analyticity characteristic. Notably, the XY model also allows for the occurrence of divergence of rate function within the same phase, with detailed analysis provided \cite{SM}.

So far, our examples have been restricted to one-dimensional systems. To further demonstrate the applicability of our approach in two-dimensional models, we now turn to the Haldane model, whose Hamiltonian is given by
\begin{equation}
\begin{split}
& H =\ t_1 \sum_{\langle i,j \rangle} \left( c_{A,\vec{r}_i}^\dagger c_{B,\vec{r}_j} + \text{H.c.} \right) \\
&+ t_2 \sum_{\langle\!\langle i,j \rangle\!\rangle} \left( e^{-i\theta} c_{A,\vec{r}_i}^\dagger c_{A,\vec{r}_j} + e^{i\theta} c_{B,\vec{r}_i}^\dagger c_{B,\vec{r}_j} + \text{H.c.} \right) \\
&+ M \sum_j \left( c_{A,\vec{r}_j}^\dagger c_{A,\vec{r}_j} - c_{B,\vec{r}_j}^\dagger c_{B,\vec{r}_j} \right).
\end{split}
\end{equation}
Here, $\langle i,j \rangle$ and $\langle\!\langle i,j \rangle\!\rangle$ denote nearest-neighbor and next-nearest-neighbor pairs, respectively. The parameters $t_1$ and $t_2$ are the corresponding hopping amplitudes, $\theta$ is the effective phase associated with next-nearest-neighbor hopping, and $M$ represents the staggered sublattice potential.

\begin{figure}[!h]
  \centering
  \includegraphics[width=8.25cm]{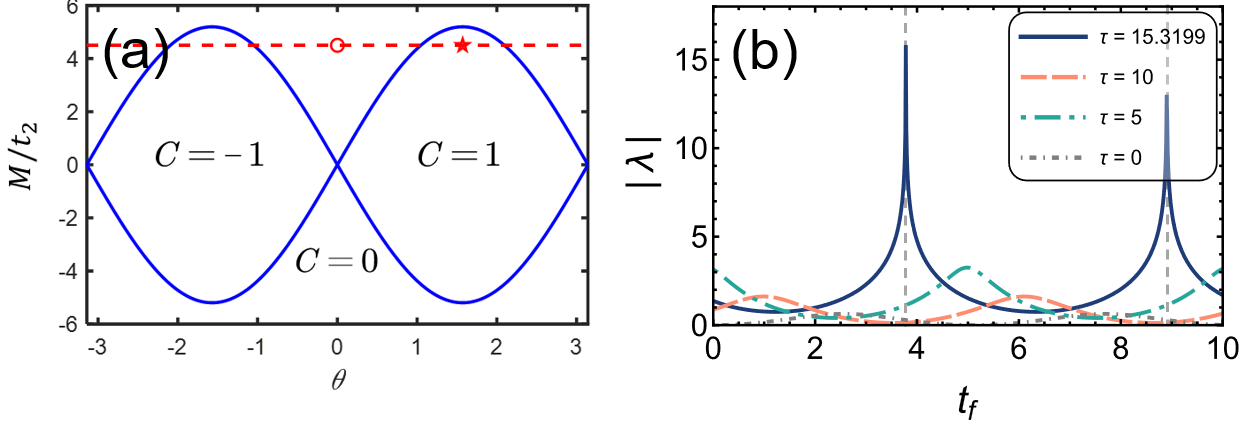}
  \caption{(a) Phase diagram of the Haldane model. The solid blue lines represent the critical lines, and the dashed red line represents parameters at $M/t_2=4.5$. The red circle represents the parameters of the initial state, and the pentagram represents the final.
  (b) Time evolution of the rate function for Haldane model.}
  \label{fighaldane}
\end{figure}

For a finite system with $L_x=L_y=50$, we choose parameters $t_1=t_2=1$, $M=4.5$, $\theta_i=0$, and $\theta_f=\tfrac{\pi}{2}$, as shown in Fig.\ref{fighaldane}(a). To illustrate  LEZs in the finite-size Haldane model, we select the effective momentum $\boldsymbol{k}=\left(\tfrac{4\sqrt{3}\pi}{15}, \tfrac{2\pi}{3}\right)$. In this case, the quench rate $\tau=15.3199$ yields a strict divergence of the rate function, as shown in Fig.\ref{fighaldane}(b).

{\it Conclusion.-}
We have proposed a scheme for seeking exact LEZs in a finite quantum Ising chain by using a two-step quenching protocol.  By analyzing the condition for the  occurrence of LEZs, we unveil that exact LEZs can be always achieved by tuning the quench rate when the systems are driven across the phase transition point. The existence of LEZs leads to the divergence of rate function of a finite quantum system at a series of critical times. Decreasing the quench rate increases the time intervals between the occurrence of divergence of rate function, and the finite size gives a lower bound of quench rate, below which no LEZ occurs.

Our work provides a practical and experimentally feasible framework for exploring LEZs  in finite-size quantum systems with controlled quenches. Our results unveil the intricate interplay between quench dynamics, system size and the underlying Hamiltonian, bridging the gap for understanding dynamical critical phenomena of quantum systems with finite size and in the thermodynamic limit.

\begin{acknowledgments}
This work is supported by National Key Research and Development Program of China (Grant No. 2021YFA1402104) 
and the NSFC under Grants  No.12474287  and No. T2121001.
\end{acknowledgments}

\section{DATA AVAILABILITY}
The data that support the findings of this article are openly available\cite{data}.

\end{document}